\documentclass[a4paper,conference]{IEEEtran}
\usepackage[latin9]{inputenc}
\usepackage{url}
\usepackage{amsmath}
\usepackage{amssymb}
\usepackage{graphicx}
\usepackage{esint}

\makeatletter

\DeclareTextSymbolDefault{\textquotedbl}{T1}
\providecommand{\tabularnewline}{\\}

\IEEEoverridecommandlockouts
\usepackage{cite}
\usepackage{amsfonts}
\usepackage{algorithmic}
\usepackage{textcomp}
\usepackage{xcolor}
\def\BibTeX{{\rm B\kern-.05em{\sc i\kern-.025em b}\kern-.08em
    T\kern-.1667em\lower.7ex\hbox{E}\kern-.125emX}}

\makeatother

\begin{document}
\title{Physics-informed generative neural networks  \\for RF propagation prediction with application to \\indoor body perception \thanks{This work is funded by the European Union. Views and opinions expressed are however those of the author(s) only and do not necessarily reflect those of the European Union or European Innovation Council and SMEs Executive Agency (EISMEA). Neither the European Union nor the granting authority can be held responsible for them. Grant Agreement No: 101099491.}}
\author{\IEEEauthorblockN{Federica Fieramosca\textit{$^{1}$}, Vittorio Rampa\textit{$^{2}$},
Michele D'Amico\textit{$^{1}$, }Stefano Savazzi\textit{$^{2}$}, } \IEEEauthorblockA{\textit{$^{1}$}\textit{\emph{ }}DEIB, Politecnico di Milano, Piazza Leonardo da Vinci 32, I-20133, Milano, Italy\textit{}\\
\textit{$^{2}$}\textit{\emph{ Consiglio Nazionale delle Ricerche}}\emph{,}
\textit{\emph{IEIIT}} institute, Piazza Leonardo da Vinci 32, I-20133,
Milano, Italy.\linebreak{}
 }}
\maketitle
\begin{abstract}
Electromagnetic (EM) body models designed to predict Radio-Frequency (RF) propagation are time-consuming methods which prevent their adoption in strict real-time computational imaging problems, such as human body localization and sensing. Physics-informed Generative Neural Network (GNN) models have been recently proposed to reproduce EM effects, namely to simulate or reconstruct missing data or samples by incorporating relevant EM principles and constraints. The paper discusses a Variational Auto-Encoder (VAE) model which is trained to reproduce the effects of human motions on the EM field and incorporate EM body diffraction principles. Proposed physics-informed generative neural network models are verified against both classical diffraction-based EM tools and full-wave EM body simulations.
\end{abstract}


\section{Introduction}

\label{sec:intro}

\IEEEPARstart{C}{ommunication} while sensing tools employ ambient radio signals used for wireless communications also to detect, locate, and track people that do not carry any electronic device, namely device-free~\cite{youssef-2007,wilson-2010,savazzi-2016}. Data analytic tools, such as Bayesian~\cite{kat} and machine learning approaches~\cite{palip}, can be used to extract information about body movements from the observed RF radiation field. 

Approaches proposed for solving the radio sensing problem require an approximated knowledge of a physics-informed (prior) model used to interpret the effects of human subjects on radio propagation~\cite{wang-2015}. The impact of body movements on radio signals can be interpreted using electromagnetic (EM) propagation theory considerations~\cite{krupka-1968}. To this aim, several physical and statistical models have been then proposed to exploit full wave approaches, ray tracing, moving point scattering~\cite{scatt} and diffraction theory~\cite{koutatis-2010,plouhinec-2019,rampa-2017,rampa-2022a}. On the other hand, a general EM model for the prediction of body-induced effects on propagation is still under scrutiny~\cite{hamilton-2014} while current full-wave models are also too complex to be of practical use for real-time sensing scenarios \cite{eleryan-2011}. 

Physics-informed generative modeling~\cite{generative_mag} is an emerging field in several application contexts ranging from EM field computation, imaging and inverse problems. For example, Generative Neural Networks (GNN) generate observations drawn from a distribution which reflects the complex underlying physics of the environment under study. Generative models for prediction of RF propagation find applications in communications and sensing~\cite{ss}. For example,~\cite{phyindoor} discussed a convolutional encoder-decoder structure that can be trained to reproduce the results of a ray-tracer, while the generation of body induced RF excess attenuation values to simulate diffraction effects was considered in~\cite{generation}. A ML-assisted channel modeling approach was proposed in~\cite{hybphydata} to generate site-specific mmWave channel characteristics. Proposed EM-informed GNN models can be trained with samples obtained either by EM diffraction-based body models \cite{plouhinec-2019,rampa-2017,rampa-2022a,eleryan-2011}, as also done in \cite{generation}, or using training samples obtained from EM body simulation tools, such as FEKO\textsuperscript{\textregistered} software. 

\begin{figure}
\centering \includegraphics[scale=0.41]{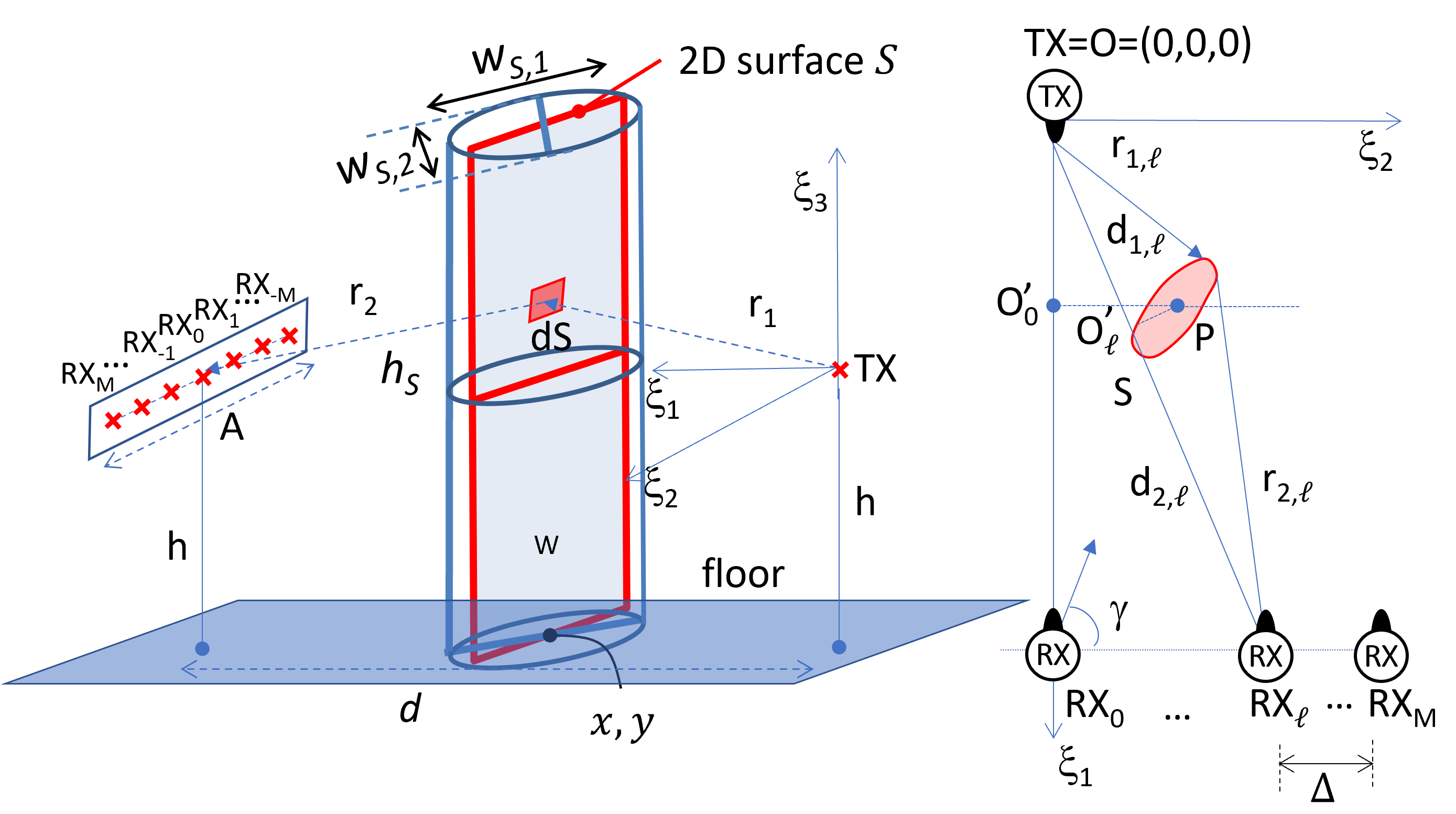} \protect\caption{\label{intro} Link geometry with body, sketched as a 2D EM perfectly absorbing sheet, and Multiple-Input-Multiple-Output (MIMO) antennas placed in the monitored area. The top view of this area is shown on the right as well.}
\end{figure}

\begin{figure}
\centering \includegraphics[scale=0.35]{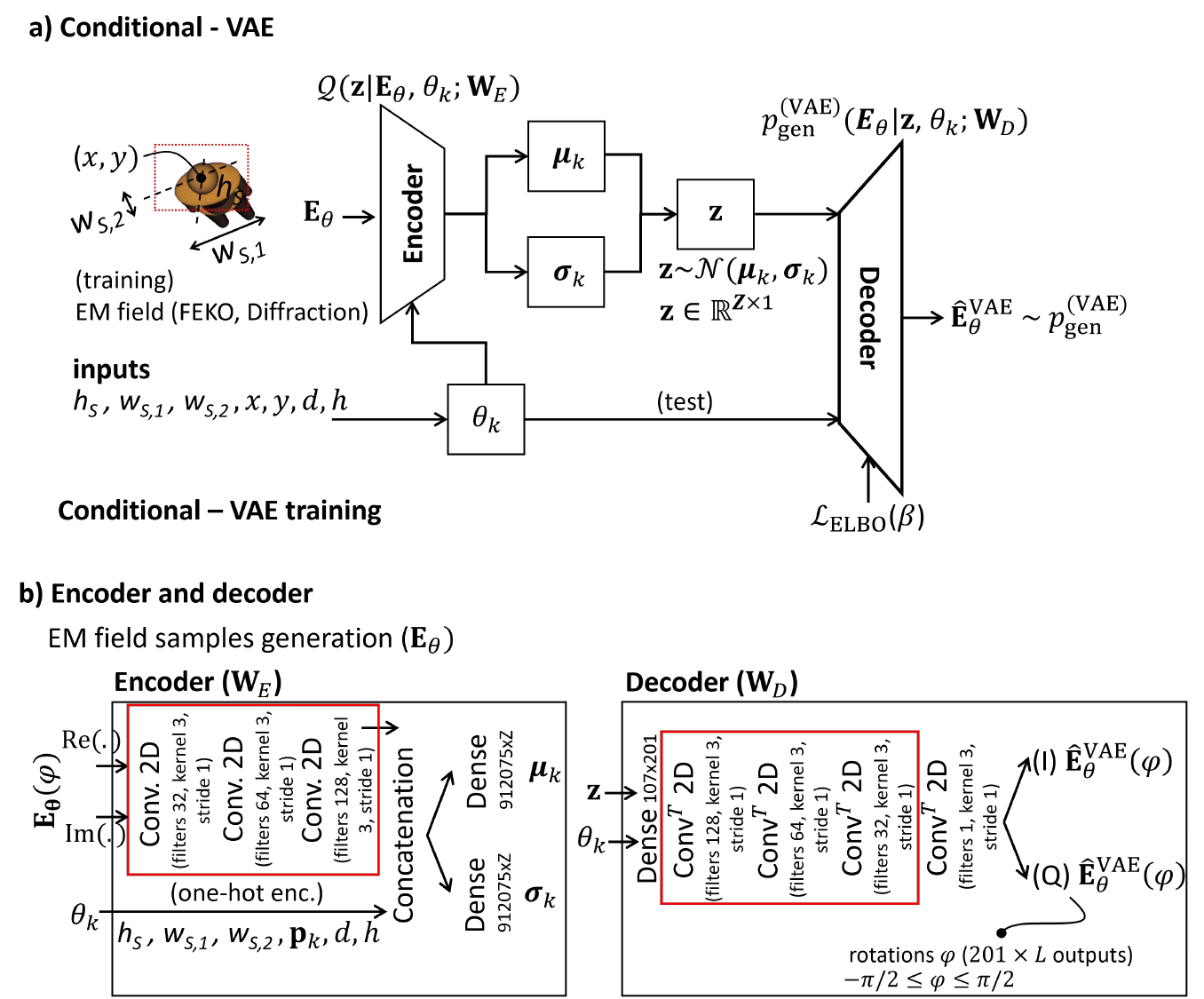} \protect\caption{\label{vae_gan} Conditional VAE system, encoder and decoder GNN structures.}
\end{figure}

\textbf{Contributions.} The paper discusses the adoption of EM-informed generative models inspired by VAE tools~\cite{bond,generation}. VAE methods can be used to build a surrogate model for EM field analysis comprising of a generator of EM field samples which is trained to reproduce EM diffraction effects \cite{rampa-2017,rampa-2022a,eleryan-2011}. The paper extends~\cite{generation} as it proposes a tool to reproduce the effects of body movements on the complex EM field, rather than the scalar body-induced RF attenuation (in fact, complex values of EM field are needed for array processing \cite{mimo}). As depicted in Fig.~\ref{intro}, the proposed generators are also optimized for multiple antenna settings, and can be used to reproduce the responses of conventional array processing. In line with initial studies~\cite{generative_mag,dibarba,baldan}, the proposed generators are purely based on sequences of convolutional (and de-convolutional) layers.

The paper is organized as follows. Sect.~\ref{sec:Generative-models-in} introduces the problem and the setup adopted for simulation, while Sect.~\ref{sec:Diffraction-models} discusses relevant EM body models for passive radio sensing with multiple antennas. Sect.~\ref{sec:EM-informed-generative-models} targets the problem of generative modelling, and discusses adaptations to VAE tools. Sect.~\ref{sec:Model-validation} verifies the models effectiveness in reproducing the array response in typical dense MIMO configurations and compared with two EM body models. Conclusions are finally discussed in Sect.~\ref{sec:conclusions}.

\section{GNNs for reproducing RF propagation effects\label{sec:Generative-models-in}}

The goal of radio sensing is to extract the EM effects ($\mathbf{E}_{\theta}$) of human body movements on propagation, based on the noisy measurements $\mathbf{S}_{t}$ of the RF radiation at time $t$. The subjects, or targets, are characterized by an unknown state $\mathbf{\theta}$ which we want to recover in real-time. Subject state might provide information about motions, location, size/shape, and orientation~\cite{kat,rampa-2017,rampa-2022a} of the bodies. As depicted in Fig. \ref{intro}, we consider a scenario where the EM body effects $\mathbf{E}_{\theta}$ are observed by a dense array of receivers. Following the classical Bayesian formulation, the objective is to maximize the a posterior distribution $p(\mathbf{E}_{\theta}|\mathbf{S}_{t})$:
\begin{equation}
p(\mathbf{E}_{\theta}|\mathbf{S}_{t})=\frac{p(\mathbf{S}_{t}|\mathcal{\mathbf{E}_{\theta}})\cdot p(\mathbf{E}_{\theta})}{p(\mathbf{S}_{t})}\label{eq:post}
\end{equation}
of the (unknown) effects $\mathbf{E}_{\theta}$, given the measurements $\mathbf{S}_{t}$. Measurements considered here are in the form of complex base-band channel impulse response (CIR)~\cite{savazzi-2016} obtained from radio devices equipped with multiple antenna frontends. The observations $\mathbf{S}_{t}$ are perturbed by body movements according to the prior distribution $p(\mathbf{E}_{\theta})$, which models the expected effects of the body in the state $\mathbf{\theta}$ as the result of the RF propagation.

Solving the radio sensing problem requires the knowledge of the RF measurements model $p(\mathbf{S}_{t}|\mathbf{E}_{\theta})$, and the prior distribution $p(\mathbf{E}_{\theta})$. While the first term depends on the data collection process, the prior distribution $p(\mathcal{\mathbf{E}_{\theta}})$ is usually hard to model as it often requires time-consuming full-wave EM approaches. Therefore, these methods are commonly considered of limited practical use for real-time sensing scenarios~\cite{eleryan-2011}. In line with~\cite{generation}, the proposed GNN model approach is designed to generate samples from the prior distribution $p(\mathcal{\mathbf{E}_{\theta}})$. Generation times are analyzed to assess the feasibility of using the proposed model during real-time target tracking as well. 

\section{EM body modelling with multiple antennas \label{sec:Diffraction-models}}

In this section, we discuss relevant EM body models to reproduce the EM effects $\mathbf{E}_{\theta}$ of body movements. As far as the diffraction-based model is concerned, we adopted the body model originally proposed in~\cite{rampa-2017} for a single link scenario and extended in \cite{mimo} for the case of multiple array receivers. 

Full-wave electromagnetic simulations were performed with Altair\textsuperscript{\textregistered} FEKO\textsuperscript{\textregistered}, a commercial software package that implements, among others, the Method of Moments (MoM), a full-wave solution of Maxwell's integral equations in the frequency domain. These calculations are computationally-intensive: in fact, as described in Sect.~\ref{sec:Model-validation}, the surface of the humanoid obstacle is discretised using triangles, then the Surface Equivalent Principle (SEP) is applied by introducing equivalent electric and magnetic currents on the surface of the dielectric body. 

Besides full-wave calculations, that are needed to represent the EM field, most of physical models designed to quantify realistic body motions must also take into account imperfect knowledge of the environment, namely small and involuntary body movements~\cite{rampa-2017} which cause the subject to change its orientation or pose w.r.t. to the receivers. All these impairments make body features hard to obtain with an acceptable level of accuracy. A statistical approach is often adopted, being in line with classical Bayesian informative-prior modeling. We assume that EM effects $\mathbf{E}_{\theta}$ are obtained for random instances of body configuration/features $\theta$, which follow the probability function $p(\theta|\theta_{k})$. Formally, the received EM field $\mathbf{E}_{\theta}$ is thus defined in probability as it is sampled from a distribution $p(\mathbf{E}_{\theta})$ defined as: 
\begin{equation}
p(\mathbf{E}_{\theta}|\theta_{k})=\mathbf{\mathbf{E}}_{\theta\sim p(\theta|\theta_{k})},\label{eq:samp}
\end{equation}
where the function $p(\theta|\theta_{k})$ models the uncertainty with respect to the \emph{nominal body features} $\theta_{k}$.

\section{Physics-informed generative neural network \label{sec:EM-informed-generative-models}}

The generative models considered here are able to reproduce different body-induced EM effects $\mathbf{E}_{\theta}$ as sampled from the ground-truth prior distribution $p(\mathbf{E}_{\theta}|\theta_{k})$ in (\ref{eq:samp}), and conditioned on the input nominal body features, or configurations, $\theta_{k}$. Unlike~\cite{generation}, body effects are represented here by the EM field responses $\mathbf{E}_{\theta}$, while the generative tool is designed to reproduce the body effects observed for each receiving antenna/link $\ell$. We also limit our focus on simple body features $\theta_{k}=\left[\mathbf{p}_{k},\varphi_{k},h_{S},w_{S,1},w_{S,2}\right]$ that include body locations $\mathbf{p}_{k}$, relative orientations $-\pi/2\leq\varphi_{k}\leq\pi/2$, height $h_{S}$, max $w_{S,1}$ and min $w_{S,2}$ traversal sizes of the target, which could, for example, represent a change in the subject pose (see Sect.~\ref{subsec:Analysis}). 

\subsection{Conditional Variational Auto-Encoder}

As depicted in Fig.~\ref{vae_gan}, the proposed VAE model, namely conditional VAE, or C-VAE for short, uses an encoder $\mathcal{Q}(\mathbf{z}|\mathbf{E}_{\theta},\theta_{k};\mathbf{W}_{E})$, parameterized by the NN parameters $\mathbf{W}_{E}$, which learns the latent space $p_{\mathcal{Z}}(\mathbf{z}|\theta_{k})\sim\mathcal{N}(\mathbf{\boldsymbol{\mu}_{\mathit{k}}},\boldsymbol{\sigma}_{k}^{2})$ for inputs $\theta_{k}$, assumed here as multivariate Gaussian distributed, with mean $\mathbf{\boldsymbol{\mu}_{\mathit{k}}}$ and standard deviation $\boldsymbol{\sigma}_{k}$ parameters. Note that other choices are not investigated here. The encoder uses training samples of (true) body-induced values $\mathbf{E}_{\theta}$ and the corresponding features $\theta_{k}$ of interest. The decoder produces a distribution $\widehat{\mathbf{E}}_{\theta}^{\mathrm{VAE}}\sim p_{\mathrm{gen}}^{\mathrm{VAE}}(\mathbf{E}_{\theta}|\theta_{k})$
\begin{equation}
p_{\mathrm{gen}}^{\mathrm{VAE}}(\mathbf{E}_{\theta}|\theta_{k})=\int_{\mathcal{Z}}p_{\mathrm{gen}}^{\mathrm{VAE}}(\mathbf{E}_{\theta}|\mathbf{z},\theta_{k};\mathbf{W}_{D})\,p_{\mathcal{Z}}(\mathbf{z}|\theta_{k})\,d\mathbf{z},\label{eq:vaegen}
\end{equation}
which is the marginalization of the conditional probability $p_{\mathrm{gen}}^{\mathrm{VAE}}(\mathbf{E}_{\theta}|\mathbf{z},\theta_{k};\mathbf{W}_{D})$ function of the NN parameters $\mathbf{W}_{D}$. The goal is to maximize the likelihood bound namely Evidence Lower BOund (ELBO) $\mathcal{L}_{\mathrm{ELBO}}$ as described in~\cite{vae-1}.

\begin{table}[tp]
\protect\caption{\label{generation_times}C-VAE vs diffraction model and FEKO\textsuperscript{\textregistered} simulations: model configuration and EM field generation time analysis.}
\vspace{0.3cm}
 
\begin{centering}
\begin{tabular}{|l|l|l|}
\hline 
\multicolumn{1}{|c|}{\textbf{Model}} & \multicolumn{1}{c|}{\textbf{Configuration}} & \multicolumn{1}{c|}{\textbf{Generation time}}\tabularnewline
 & ($L$ links) & {[}s/sample/link{]}\tabularnewline
\hline 
 & $Z=16$, $L=1$ & $4.6\times10^{-5}$ \tabularnewline
\textbf{C-VAE} & $Z=32$, $L=1$  & $6.3\times10^{-5}$\tabularnewline
\textbf{$(\widehat{\mathbf{E}}_{\theta}$ }generation\textbf{)} & $Z=16$, $L=81$  & $5.3\times10^{-2}$\tabularnewline
 & $Z=32$, $L=81$ & $6.1\times10^{-2}$\tabularnewline
\hline 
 & $\epsilon=10^{-3}$ , $L=1$ & $5.4\times10^{-3}$ \tabularnewline
\textbf{EM Diffraction}  & $\epsilon=10^{-6}$, $L=1$ & $3.81\times10^{-2}$ \tabularnewline
(omnidirectional antennas)  & $\epsilon=10^{-3}$, $L=81$ & $0.78$\tabularnewline
 & $\epsilon=10^{-6}$, $L=81$ & $3.1$\tabularnewline
\hline 
\textbf{EM FEKO\textsuperscript{\textregistered} simulations} & $L=81$ & $>240$ 
\tabularnewline
\hline 
\end{tabular}
\par\end{centering}
\medskip{}
 \vspace{-0.6cm}
 
\end{table}

\subsection{Implementation considerations}

C-VAE pre-trained models reproducing diffraction effects are available on-line~\cite{github} together with example codes for generating EM field samples ($\widehat{\mathbf{E}}_{\theta}$) according to specific body configurations. Generation times of C-VAE are compared in Tab.~\ref{generation_times} with those obtained by classical diffraction models~\cite{rampa-2022a} as well as popular EM body simulator tools (Altair\textsuperscript{\textregistered} FEKO\textsuperscript{\textregistered} software, see Sect.~\ref{sec:Model-validation}). In the example, samples are generated assuming $L=81$ MIMO links with variable number of antennas at TX and RX. VAE-based generation of the samples is about one order of magnitude faster than EM diffraction model computation, which also depends on the chosen numerical integration configurations (i.e., tiled integration method, and absolute error tolerance $\epsilon$), target size, and antenna configuration. The generative model can produce up to $50\div100$ samples per second, which is reasonable considering typical body movement speeds (max. $1$m/s). 

\section{Generation of EM body model samples \label{sec:Model-validation}}

C-VAE sample generation is assessed here at carrier frequency $f_{c}=2.4$ GHz. First, we analyze the reproduction accuracy against EM body diffraction~\cite{rampa-2022a}. Next, we compare the results with FEKO\textsuperscript{\textregistered} simulations. The RX node is equipped with a Uniform Linear Array (ULA) of $9$ omni-directional antennas and spaced at $\triangle=\lambda/2$. The TX node is equipped with a single antenna while $L=9$ links are considered. All the links of the array are horizontally placed at height $h=0.99$ m from the ground. Distance between the TX and the central RX of the array is $d=4$ m. C-VAE is trained here using examples of EM diffraction from: 1) $75$ marked locations inside the first Fresnel ellipsoid of the central link of the array; 2) varying target dimension $h_{S}$ in the interval $[1.65 - 2.00]$ m.

\begin{figure}
\centering \includegraphics[scale=0.35]{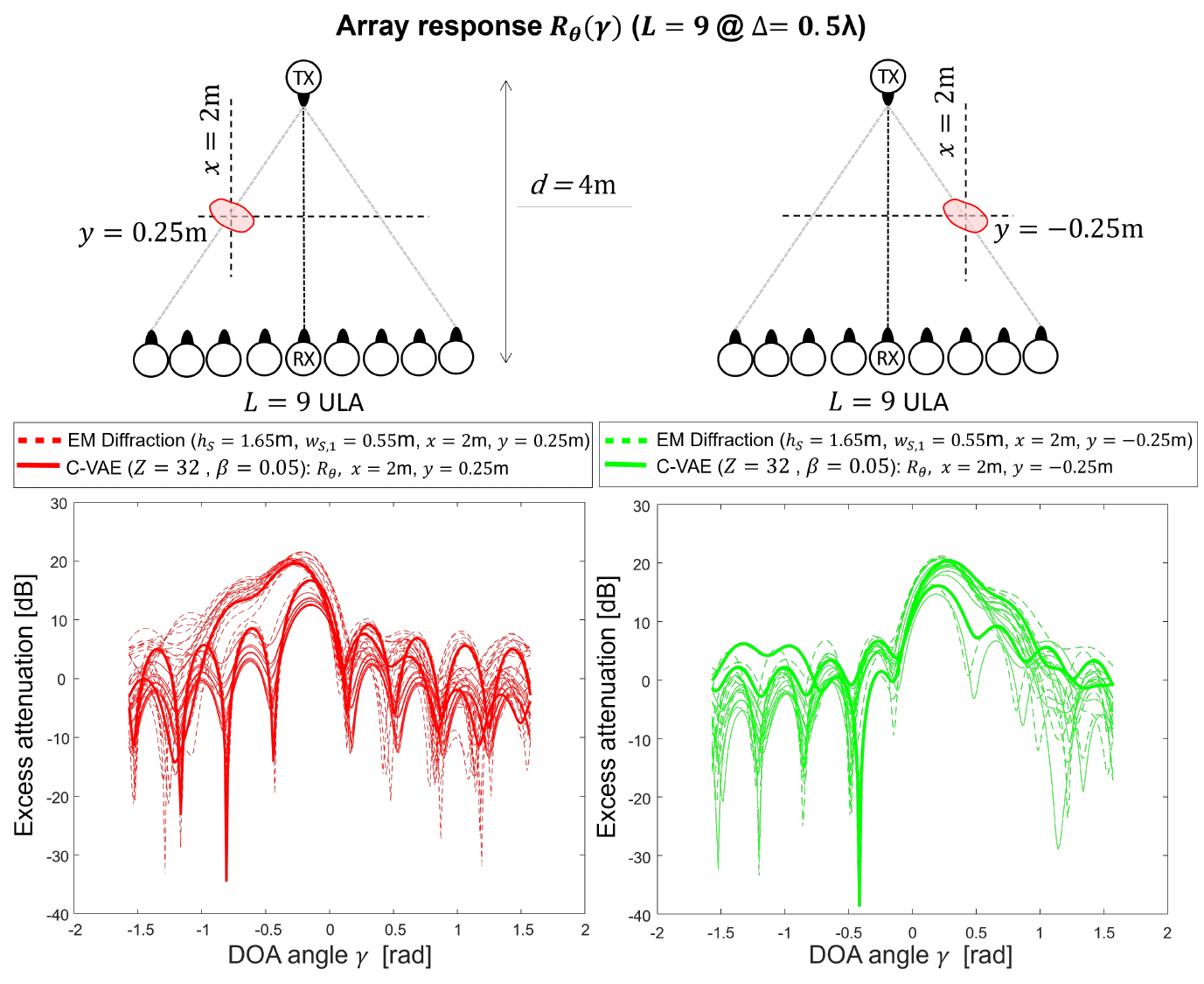} \protect\caption{\label{response} Array layout setups (top) and corresponding responses (bottom) $R_{\theta}(\gamma)$ for $\gamma=\pi/2$. Target has dimensions $h_{S}=1.65$ m, $w_{S,1}=0.55$ m, $w_{S,2}=0.25$ m. Two subject positions are considered, namely $x=2$ m, $y=0.25$ m (left) and $x=2$ m, $y=-0.25$ m (right). Response is obtained using VAE-generated field samples $\widehat{\mathbf{E}}_{\theta}^{\mathrm{VAE}}$ (green/red solid lines) and compared with the array response obtained with the EM body diffraction model (dashed lines).}
\end{figure}

In the following, we verify the ability of the proposed C-VAE model to generate samples of the EM field $\widehat{\mathcal{\mathbf{E}}}_{\theta}^{\mathrm{VAE}}=\left[\widehat{E}_{\ell,\theta}^{\mathrm{VAE}}\right]{}_{\ell=1}^{L}$ which are able to reproduce the array excess attenuation response of the conventional linear beamforming processing \cite{mimo,benesty-2021}. For $L=2M+1$ antennas ($M=4$), we consider the vector $\mathbf{w}(\gamma)=\left[w_{k}\right]_{k=-M,...,0,...,+M}=\left[w_{-M}\,...\,w_{-1}\,w_{0}\,w_{1}\,...\,w_{M}\right]^{T}$ of linear beamforming coefficients designed to \emph{steer} the antenna array in a direction $\gamma$. The received base-band signal $r_{\theta}(\gamma)$ at the output of the beamforming processing is:
\begin{equation}
r_{\theta}(\gamma)=\sum_{m=-M}^{+M}w_{m}^{*}(\gamma)\,\left[E_{\ell,\theta}+n_{\ell}\right]=\mathbf{w(\gamma)}^{H}\cdot\left[\mathbf{E}_{\theta}+\mathbf{n}\right],\label{eq:beam_forming-1}
\end{equation}
where $n_{\ell}$ is the Additive White Gaussian Noise (AWGN) complex vector $\mathbf{n}=\left[n_{-M}\,...n_{-1}\,n_{0}\,n_{1}\,...\,n_{M}\right]^{T}$ of size $2M+1$, with covariance $\sigma^{2}\mathbf{I}$. We define the body-induced excess attenuation response $F_{\theta}(\gamma)$ with target in state $\theta$ as 
\begin{equation} 
F_{\theta}(\gamma|\mathbf{E}_{\theta})=\sum_{m=-M}^{+M}w_{m}^{*}(\gamma)\,\frac{E_{\ell,0}}{E_{\ell,\theta}}=\mathbf{w(\gamma)}^{H}\cdot\frac{\mathbf{E}_{0}}{\mathbf{E}_{\theta}}.\label{eq:beam_forming-2}
\end{equation}
with $\mathbf{E}_{0}$ being the EM field measured in free space and $\gamma$ the DoA (Direction Of Arrival), namely the direction of propagation of the impinging wavefront w.r.t. the array axis.
 
\subsection{Generation of antenna array responses}
\begin{figure}
\centering \includegraphics[scale=0.37]{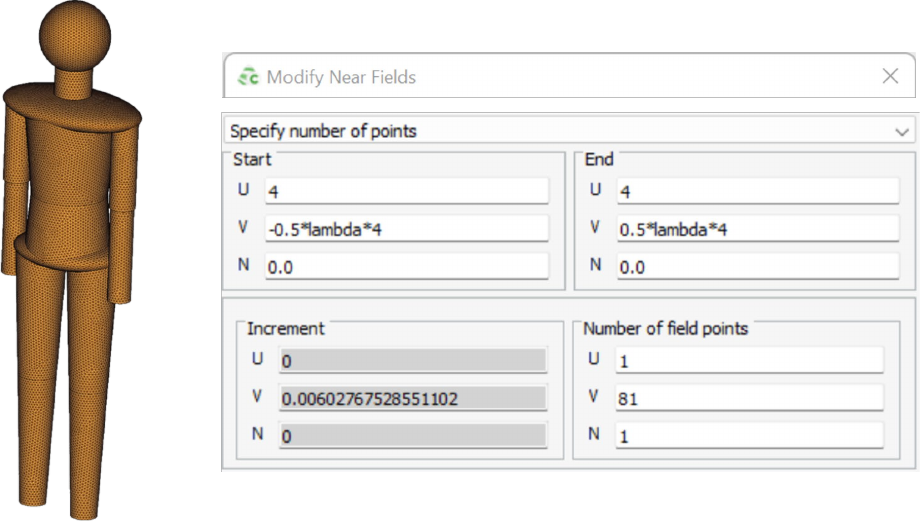} \protect\caption{\label{anto} Body-shape model built in FEKO\textsuperscript{\textregistered} with size $h_{S}=1.80$ m, $w_{S,1}=0.52$ m, and $w_{S,2}=0.32$ m. Simulation settings are shown as well.}
\end{figure}

\begin{figure}
\centering \includegraphics[scale=0.44]{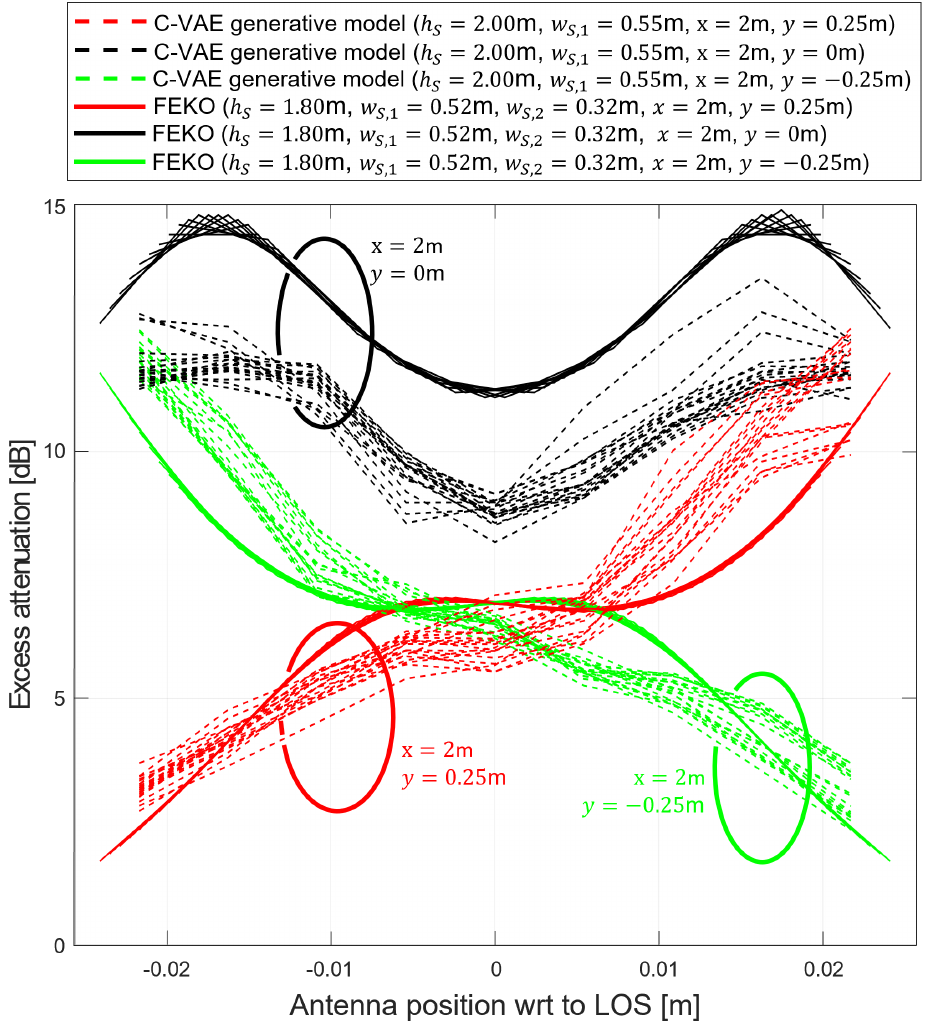} \protect\caption{\label{att} Excess attenuation values obtained by FEKO\textsuperscript{\textregistered} software (solid) and compared with C-VAE generation (dashed) for varying antennas of the array (9 antennas) and target dimensions. Three nominal subject positions are considered, namely $x=2$ m, $y=0.25$ m (red), $x=2$ m, $y=0$ m (black) and $x=2$ m, $y=-0.25$ m (green). For all cases, attenuation responses account for  small body movements around the nominal positions.}
\vspace{-0.2cm}
\end{figure}

\begin{figure}
\centering \includegraphics[scale=0.5]{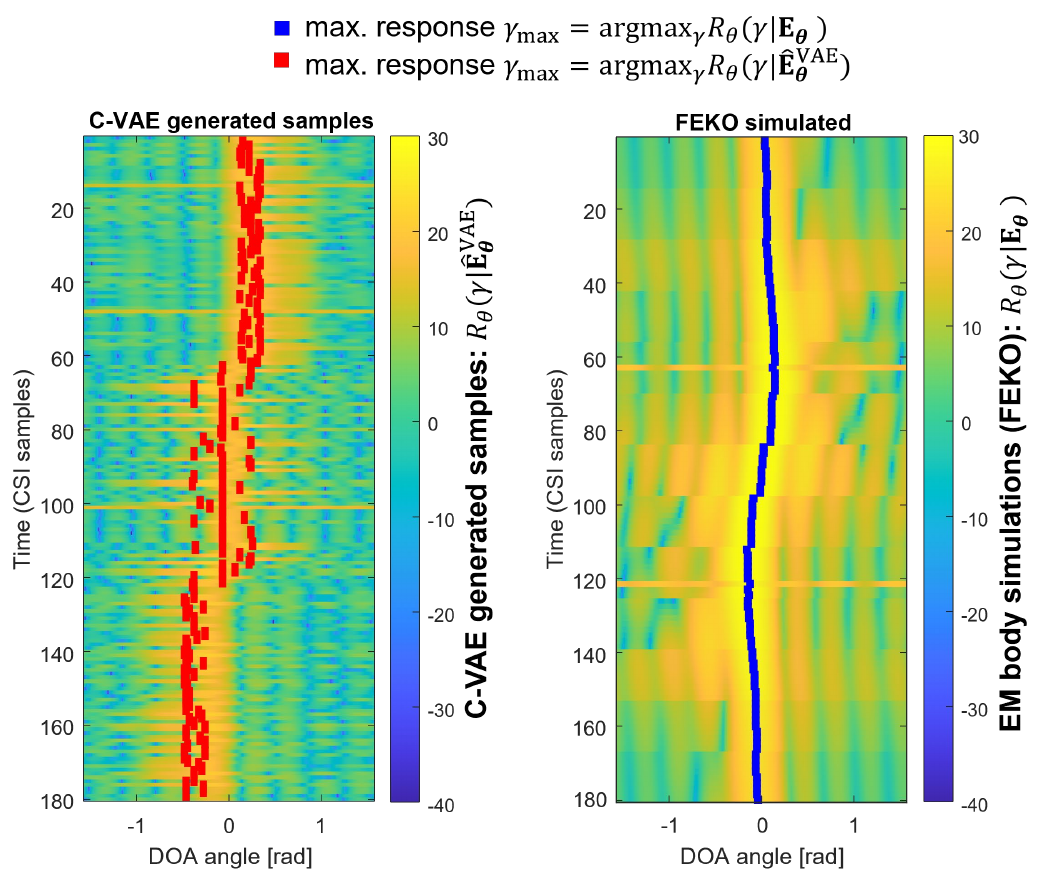} \protect\caption{\label{response-1} DoA analysis for a subject moving across the LOS ($-0.25$ m $\leq y$ $\leq0.25$ m, $x=2$ m). Comparative analysis of array responses obtained from C-VAE generation and FEKO\textsuperscript{\textregistered} simulations.}
\end{figure}

The C-VAE model is set here to reproduce the response obtained by EM body diffraction $F_{\theta}(\gamma|\widehat{\mathbf{E}}_{\theta}^{\mathrm{VAE}})\sim F_{\theta}(\gamma|\mathbf{E}_{\theta})$ through the generation of the corresponding EM field samples $\widehat{\mathbf{E}}_{\theta}^{\mathrm{VAE}}$. For conventional ULA scenarios, that assume planar wavefront propagation, the steering vector $\mathbf{w}(\gamma)=\left[a_{m}\right]$ for the array is given by eq. (7) of~\cite{mimo} (see also~\cite{benesty-2021} for details).


Fig.~\ref{response} shows the body-induced array response, in terms of the excess attenuation $20\,\log_{10}|F_{\theta}(\gamma)|$ as a function of the DoA $\gamma$ and for different values of the $y$ displacement of the target (w.r.t. the central LOS) and $x=2$ m. The human target has height $h_{S}=1.65$ m, traversal max. and min. sizes equal to $w_{S,1}=0.55$ m and $w_{S,2}=0.25$ m, respectively. The array signal processing extracts the response for varying DoA $\gamma$ and is based on Fast Fourier Transform (FFT) with $N_{FFT}=257$ points. The figure shows two responses (red and green lines) for corresponding nominal displacements $y=0.25$ m and $y=-0.25$ m, respectively. Dashed lines show the responses $F_{\theta}(\gamma|\mathbf{E}_{\theta})$ obtained using EM body diffraction, while solid lines represent the reproduced response $F_{\theta}(\gamma|\widehat{\mathbf{E}}_{\theta}^{\mathrm{VAE}})$ using C-VAE. Notice that, in the example, each array response represents a random movement of the body in the surrounding of the nominal displacement. The maximum response of the array, namely the dominant DoA $\gamma_{\mathrm{max}}=\arg\max_{\gamma}F_{\theta}(\gamma|\mathbf{E}_{\theta})$, is perturbed by the presence of the subject and such alteration is generally well reproduced by the C-VAE model. 

\subsection{Analysis with EM body simulations}
\label{subsec:Analysis}

In this section, we verify the ability of the C-VAE model to reproduce the results of EM field simulation obtained by diffraction and MoM methods (i.e., using Altair\textsuperscript{\textregistered} FEKO\textsuperscript{\textregistered} tools). The setup is again composed by a transmitter (TX) and a receiver array (RX) at $4$ m distance. The former is an electric dipole transmitting at $f=2.4868$ GHz, the latter is composed by 9 isotropic elements spaced at $\lambda/2$. For FEKO\textsuperscript{\textregistered} simulations, rather than a 2D perfectly absorbing sheet, as assumed in the scalar diffraction models~\cite{rampa-2022a}, an anthropomorphic obstacle is introduced between the two devices. The obstacle is built as shown in Fig.~\ref{anto}: a body-shape model, $h_{S}=1.80$ m, $w_{S,1}=0.52$ m and $w_{S,2}=0.32$ m, is used, which is an improvement over the 2D PEC object considered in~\cite{rampa-2017,rampa-2022a,generation,fieramosca-2023,fieramosca-2024,mimo}. The material of the body is retrieved from a simulator available online~\cite{material}. Assuming the body mainly constituted by muscles, we set: the relative permittivity $\epsilon_r= 60$, the dielectric loss tangent tan$\delta = 0.242$, and the mass density $\rho=1040.0$ kg/m$^3$. Simulations are taken with the body located in three positions: $1$) $x=2$ m and $y=0$ m (LOS, black); $2$) $x=2$ m, and $y=0.25$ m (red); $3$) $x=2$ m, and $y=-0.25$ m (green). 

In Fig.~\ref{att}, we report the body-induced excess attenuation values measured by each antenna element for the corresponding subject positions (black, red and green colors). Responses consider the effect of small body movements ($5$ cm) around the nominal subject position. FEKO\textsuperscript{\textregistered} simulated results are shown in solid lines, while in dashed lines we reported the corresponding C-VAE generated attenuation samples reproducing  EM body diffraction effects. In general, the GNN mostly underestimates the FEKO\textsuperscript{\textregistered} results, due to different hypotheses about body composition and shape, namely about $4$ dB for positions on the LOS and $2$ dB for positions outside ($y=0.25$ m, $y=-0.25$ m). A possible remedy, to be investigated in the future, is to train the GNN network using examples from both diffraction model and FEKO\textsuperscript{\textregistered} simulation samples, with the goal of improving the GNN generalization capabilities.

Fig.~\ref{response-1} compares the array responses and the DoA $\gamma_{\mathrm{max}}$ obtained from the C-VAE generator $\gamma_{\mathrm{max}}=\arg\max_{\gamma}F_{\theta}(\gamma|\widehat{\mathbf{E}}_{\theta}^{\mathrm{VAE}})$ (on the left) and FEKO\textsuperscript{\textregistered} simulated array response under the same configuration (right figure). In all cases the target is moving across the LOS ($-0.25$ m $\leq y\leq0.25$ m, $x=2$ m), with speed $0.5$ m/s. The movements of the target are distinguishable, as associated with separable DoAs, considering both generated samples (reproducing diffraction effects) and FEKO\textsuperscript{\textregistered} simulations.

\section{Concluding remarks}
\label{sec:conclusions}
The paper considered the adoption of an EM-informed Generative Neural Network (GNN) model to predict the effects of body movements on radio propagation, with applications to passive radio localization and sensing. A variational auto-encoder (VAE) tool is designed to generate samples of EM field to reproduce body-induced EM diffraction effects, considering small and involuntary movements of the body, as well as multiple antenna settings. Moreover, the C-VAE is several orders of magnitude faster than diffraction and full-wave methods. The paper verified the model effectiveness in reproducing the array response in typical MIMO configurations and the perturbations induced by body movements on such responses. Results are compared with scalar diffraction models as well as EM simulations based on FEKO\textsuperscript{\textregistered} software.

\end{document}